\documentclass[12pt]{iopart}

\usepackage{iopams}  
\usepackage{graphicx}

\renewcommand{\Im}{\mathop{\mathrm{Im}}\nolimits}

\newcommand{\eqref}[1]{\eref{#1}}

\newcount\timehh  \newcount\timemm
\timehh=\time \divide\timehh by 60
\timemm=\time
\begin{document}
\title[Collective effects in emission of quantum dots]{Collective effects in
emission of quantum dots strongly coupled to a microcavity photon}
\author{A.N. Poddubny, M.M. Glazov, and N.S. Averkiev}

\address{Ioffe Physical-Technical Institute, 26 Polytekhnicheskaya st., St.
Petersburg 194021, Russia}
\ead{poddubny@coherent.ioffe.ru}

\begin{abstract}
 A theory of non-linear emission of quantum dot ensembles coupled to the optical mode of the microcavity
is presented. Numerical results are compared with analytical approaches. The effects of  exciton-exciton interaction within the quantum dots and with the reservoir formed by nonresonant pumping are considered.  It is demonstrated, that the nonlinearity due to the interaction strongly affects the shape of the emission spectra. The collective superradiant mode of the excitons is shown to be stable against the non-linear effects.
\end{abstract}

\submitto{\NJP}
\maketitle

\section{Introduction}

Semiconductor quantum dots are often referred to as ``artificial
atoms'' owing to their discrete energy spectrum. The progress in
nanotechnology has made it possible to employ quantum dots as a
\emph{sui generis} solid state laboratory for studies of quantum
mechanics~\cite{Ivchenko2005}. Interband optical pumping of quantum dots gives
rise to the electron-hole pairs or zero-dimensional  excitons, which,
as shown in the pioneering
works~\cite{Reithmaier2004,Khitrova2004,Peter2005}, can strongly couple
with the photon trapped in the optical microcavity. The strong coupling
effect  results in a coherent energy
transfer between the photon and the exciton, this phenomenon is widely studied for bulk materials and planar quantum well 
structures~\cite{agr_ginz,kavmal2003,kavbamalas,Excitons}.
 Its observation in zero-dimensional systems has
attracted an enormous excitement of the research community since the concepts of quantum electrodynamics were directly transferred to the
solid state.
Such quantum-dot-in-a-cavity systems demonstrate fascinating fundamental
physics~\cite{Khitrova2006,Kasp2010,nomura2010,calic2011} and may be advantageous for quantum optics
device applications~\cite{Dousse2010,nomura2010,arakawa2011,tanda2011}. 

The physical concept of the strong coupling in zero-dimensional
microcavities can be easily understood considering, for the sake of an
example, two classical pendulums with close frequencies of oscillations
$\omega_1$ and $\omega_2$ connected by a spring, which induces the
coupling between the pendulums of a strength $g$. One of these
oscillators represents a photon, the other one stands for an exciton,
and the spring describes radiative recombination of the exciton into the
photon mode. If the dampings of
the individual oscillators are small compared with the coupling constant,
the eigenfrequencies $\omega$ of this interacting system follow from
the simple equation 
\begin{equation}
\label{system}
(\omega - \omega_1)(\omega-\omega_2)=g^2
\end{equation}
as
\begin{equation}
\label{frqs}
\omega_{\pm} = \frac{\omega_1+\omega_2}{2} \pm
\sqrt{\left(\frac{\omega_1-\omega_2}{2}\right)^2 +g^2}.
\end{equation}
The normal modes correspond to the coupled oscillations: By exciting
one pendulum one eventually excites another, so the energy is
transferred back and forth between them. 

The analogy between the quantum electrodynamical problem of the 
quantum dot exciton interacting with the microcavity photon and purely classical
problem of two coupled pendulums is quite deep. The excitonic
polarization in a semiconductor is indeed described by the
oscillator-like equation of motion and so does the electric field of
the cavity 
photon~\cite{Ivchenko2005,kavbamalas}. Hence, the physics of these
two different systems is similar which greatly simplifies theoretical
description of the quantum-dot-in-a-cavity
dynamics~\cite{keldysh07,JETP2009}. The situation becomes particularly interesting if
$N>1$ dots are placed in the microcavity: It turns out that,
provided the dots are identical, only one excitonic mode -- termed
\emph{superradiant} -- interacts
with the cavity mode, and the interaction constant is enhanced compared with that in one dot by the
factor $\sqrt{N}$. In classical language, for $N$ pendulums
describing quantum dot excitons, one oscillation mode is
specific, namely, the mode where all pendulums oscillate in phase
with each other.

Just like the classical oscillators which are ideal only as far as
model situations are considered, the excitons and photons interact
with the environment, which gives rise to their damping and dephasing,
moreover, the oscillation law can differ from the harmonic
one. Main reasons of unharmonicity are identified: interactions
between excitons lead to their energy shifts, the oscillator
strength saturation results in the decrease of the coupling constant
with an increase of the exciton number in the system, excitons can
bind together to form biexcitons~\cite{Ivchenko2005,jahnke}. Hence, the analysis of the non-linear dynamics of
the quantum dot excitons strongly coupled to the cavity mode is of prime
importance. In particular, the stability of the superradiant mode with
respect to the interactions should be investigated.

In this paper, we consider the simplest possible and physically most
transparent case of the non-linearity caused by the  exciton-exciton interaction. It is analogous to the cubic unharmonicity of the pendulum.\footnote{The effects of the oscillator strength saturation related with the two-level nature of exciton transition in quantum dot were considered in Refs.~\cite{Laussy2006,laussy2009,Poddubny2010prb}.}
We calculate the optical emission spectra of the quantum-dot-in-a-cavity system under a non-resonant excitation (photoluminescence
spectra). The non-resonant pumping is modeled as a random
force acting on the corresponding oscillator~\cite{JETP2009,Voronov2007}.
The parameter which controls the non-linearity is the
pumping rate: the higher the pumping, the larger the fluctuations of
excitonic polarization and, correspondingly, the higher are the
non-linear terms in the equations of motion.

In our approach, the quantum dot state is described as a classical unharmonic oscillator. Hence, this approach is valid for  large enough quantum dots where excitons are quantized as a whole.  Similar behavior can be expected in a variety of systems including the semiconductor systems with planar quantum microcavities where excitons can be trapped by the disorder and the non-linear regime is easily reached~\cite{LeSiDang06,PhysRevLett.98.206402}, or quantum well structures with dipolar excitons~\cite{butov_bec,timofeev:179}, as well as many other, e.g. optomechanical structures where the optical mode of the cavity
interacts with a classical oscillator~\cite{Aspelmeyer2010}.

The paper is organized as follows. Section~\ref{sec:model} outlines the model, Sec.~\ref{sec:reservoir} is devoted to the role of the reservoir fluctuations, Sec.~\ref{sec:nonlin} presents the study of the nonlinearity due to the exciton interaction. Main results are summarized in Conclusions, Sec.~\ref{sec:concl}.
\section{Model}\label{sec:model}

For distinctness we consider here a zero-dimensional microcavity where one or several quantum dots are embedded. We suppose that the energy (or frequency, we put $\hbar=1$ for brevity)  spacing between
the cavity modes is large enough to consider only one relevant photonic mode whose frequency $\omega_C$ is close enough to frequencies of optical transitions in quantum dots $\omega_{X,i}$, $i=1, \ldots, N$, where $N$ is the dot number. For simplicity, the
interaction of the cavity mode with the ground states of quantum dot
excitons is considered only, the treatment can be generalized to allow
for the excited states. Moreover, the polarization degree of freedom of the cavity mode and spin degrees of freedom of excitons are disregarded. Under these approximations the equations of
motion for the dimensionless electric field $E$ and excitonic
polarizations $P_i$ can be written as 
\begin{eqnarray}\label{eq:main}
   &\frac{dE}{dt}=-\left(\rmi \omega_{C}+\frac{\Gamma_{C}}{2}\right)E-\rmi g
\sum\limits_i P_{i}\:,\\\nonumber
&\frac{dP_i}{dt}=-\left(\rmi
\omega_{X,i} + \frac{\Gamma_{X,i}}{2} +\rmi\alpha_{i}|P_i|^2+  {\rmi \beta_i n_R}\right)P_i
-\rmi g
E+{f}_i(t),\quad i=1\ldots N\:.
\end{eqnarray}
Here  $\Gamma_{C}$ and $\Gamma_{X,i}$ are (phenomenological) decay
rates for the cavity and excitons, respectively, $g$ is the
light-matter coupling constant (taken the same in all dots for the
sake of simplicity). Its evaluation is beyond the current
work, rigorously it can be done by solving Maxwell equations and
Schr\"{o}dinger equation for the exciton envelope functions in quantum
dots, see Refs. \cite{Andreani1999,Kaliteevski2001b,Glazov2011}. 

Equation~(\ref{eq:main}) for exciton polarization contains also
non-linear and driving terms. Former ones, proportional to the
interaction parameters $\alpha_i>0$, describe the blueshifts of excitonic
states due to interactions within the same dot. Here and further we
assume that the quantum dot size is comparable or larger than
excitonic Bohr radius to accomodate several excitons in the dot~\cite{Laussy2006}.\footnote{Another possibility is
  to consider a planar microcavity structure with a lateral potential
  confining the exciton, see e.g. Ref.~\cite{Tosi2012}.} Terms $\propto \beta_i n_R$ take into account the
interaction of quantum dot excitons with a reservoir formed, e.g. by
excitons and electron-hole pairs within the wetting layer. The
reservoir population $n_R$ is, in general, function of the pumping
intensity $W$ and exciton occupations, $n_R\equiv
n_R(W,|P_1|^2,\ldots, |P_N|^2)$. Driving terms described by random
forces $f_i$, $\langle f_i \rangle =0$ account for the exciton generation in quantum dots caused
by their relaxation from the wetting layer and excited states, see
Refs.~\cite{Voronov2007,JETP2009} for details. These random forces represent the white noise
\begin{equation}
 \langle f_i(t)f_j^*(t')\rangle=S_{i}\delta_{ij}\delta(t-t')\:,\label{eq:random}
\end{equation}
characterized by the exciton generation rate $S_i$ proportional to the reservoir population, $S_i = s_i n_R$, where $s_i$ is related with the relaxation rate of excitons towards the ground state~\cite{JETP2009}.
In what follows we assume that higher-order correlators of random
forces are reduced to the second order ones, Eq.~(\ref{eq:random}), in
accordance with the Gaussian distribution. Such random forces model
the incoherent non-resonant pumping of our system. Similar approaches
were used to study optical emission of exciton-polaritons in Bragg
multiple quantum well structures and for planar 
microcavities~\cite{Voronov2007,Wouters2009,Sarchi2009,wouters2010}.

We are ultimately interested in the luminescence spectrum given (up to
the common factor) by~\cite{JETP2009}:
\begin{equation}
\label{Lum}
I(\omega) = \langle |E(\omega)|^2 \rangle = \int_{-\infty}^\infty
\langle E(t+t')E^*(t) \rangle e^{\rmi \omega t'}\mathrm dt',
\end{equation}
where averaging over time $t$ is assumed. If the non-linear
contributions are disregarded, emission spectrum can be obtained
analytically with the result~\cite{JETP2009}:
\begin{equation}\label{eq:linear}
 I_{\rm lin}(\omega)=\sum\limits_{m,m'=1}^{N+1}\frac{C^{(m)*}_{\rm cav}C^{(m')}_{\rm cav}
\sum_{i=1}^N C^{(m)*}_iS_iC^{(m')}_i
}{(\Omega_m^*-\omega)(\Omega_{m'}-\omega)}\:.
\end{equation}
Here, $\Omega_m$ are the eigenfrequencies of the homogeneous system (\ref{eq:main}) found neglecting non-linear terms $\propto \alpha_i$, $\beta$ i.e. $\Omega_m$ are the polariton frequencies, and
$[C_{\rm cav}^{(m)},C_{1}^{(m)},...,C_{N}^{(m)}]\equiv [E,P_1,\ldots,P_N]$
are the corresponding eigenvectors, i.e., the Hopfield coefficients of the excitonic polaritons formed by the excitons coupled to the cavity mode~\cite{kavbamalas}. The spectrum Eq.~\eqref{eq:linear} can also be recast as a sum of terms with poles at polariton frequencies $\Omega_m$, $\Omega_m^*$.

The detailed analysis of Eq.~\eqref{eq:linear} is presented in Ref.~\cite{JETP2009}. Here we briefly discuss an important liming case where all exciton frequencies are the same, $\omega_{X,1} = \omega_{X,2} = \ldots = \omega_{X,N} \equiv \omega_X$, exciton decay rates are the same, $\Gamma_{X,1} = \Gamma_{X,2} = \ldots = \Gamma_{X,N} \equiv \Gamma_X$ and the decay rates are negligible compared with the coupling constants, $\Gamma_X, \Gamma_C \ll g$. Under these assumptions the polariton frequencies and Hopfield coefficients take simple form. There are two mixed modes with the frequencies, cf. Eq.~\eqref{frqs},
\begin{equation}
\label{super}
\Omega_{1,2} = \frac{\omega_X + \omega_C}{2} - \mathrm i \frac{\Gamma_X+ \Gamma_C}{4} \pm \sqrt{ \left(\frac{\omega_X - \omega_C}{2}\right)^2 + N g^2},
\end{equation}
and all the remaining $N-1$ modes correspond to the exciton states decoupled from light, $\Omega_m=\omega_X$ for $m=3,\ldots N+1$. The effective coupling strength is enhanced by the factor $\sqrt{N}$ due to the superradiance effect: dipole moments of excitons oscillate in phase. Correspondingly, emission spectrum has two peaks at $\Omega_1$ and $\Omega_2$ splitted by $\sqrt{N}g$:
\begin{equation}
I_{\rm lin}(\omega)\equiv\langle |E(\omega) |^2\rangle\propto
\frac1{|(\omega-\omega_X+\rmi\Gamma_X/2)(\omega-\omega_C+\rmi\Gamma_C/2)-Ng^2|^2}
\:.\label{eq:PLlin}
\end{equation}
We note that Eq.~\eqref{eq:PLlin} is valid even if the condition $\Gamma_X \ll g$ or $\Gamma_C \ll g$ is violated~\cite{JETP2009}. Two distinct peaks are observed in the emission spectra for the broadenings as high as $\Gamma_X,\Gamma_C \sim \sqrt{N} g$, otherwise these two peaks merge into one.

Now we turn to the discussion of the non-linear effects. Firstly, we
consider interactions with the reservoir described by the terms
$\propto \beta_i n_R P_i$, afterwards we discuss the effects of
exciton-exciton interaction within the same dot described by the terms
$\propto \alpha_i |P_i|^2P_i$.

\section{Interaction with reservoir}\label{sec:reservoir}
Under conventional non-resonant pumping conditions the excitons are
generated in the wetting layer and form a reservoir. If the pumping
rate is moderate the majority of excitons are in the reservoir and its occupation
is only weakly affected by the presence of the quantum dots, hence,
one can treat
$n_R$ in Eqs.~(\ref{eq:main}) as an independent
quantity. For the steady-state pumping $\langle n_R \rangle = \bar
n_R$ and there are certain fluctuations of $\delta n_R$ around this
time-averaged value, $\delta n_R = n_R - \bar n_R$.

To elucidate the role  of the reservoir we neglect completely the
non-linearities caused by the exciton-exciton interaction in quantum dots,
i.e. we put $\alpha_i\equiv 0$ in Eqs.~(\ref{eq:main}). It is
instructive to analyze the case of a single quantum dot and
disregard the light-matter interaction. In this situation we obtain
\begin{equation}
\label{1:res}
\frac{dP}{dt}=-\left(\rmi
\omega_{X} + \frac{\Gamma_{X}}{2}  + \rmi \beta \bar n_R + \rmi \beta
\delta n_R\right)P +f(t).
\end{equation}
For clarity we have separated in Eq.~(\ref{1:res}) the contributions
due to mean number of particles in reservoir and due to its
fluctuations. Equation~(\ref{1:res}) is the first order linear
differential equation which can be integrated with the result
\[
P(t) = \int_0^t \mathrm dt' f(t') \exp{\left[-(\mathrm i \omega_X + \mathrm i \beta \bar n_R + \Gamma_X/2 )(t-t') - \mathrm i \beta \int_{t'}^t \delta n_R(t'')\rmd t'' \right]},
\]
where the solution of homogeneous equation is omitted. The autocorrelation function $\langle P(t) P^*(0)\rangle$, whose Fourier transform determines the single dot emission spectrum in the case of the regime of the weak coupling with the photon, reads
\begin{equation}
\label{1:res:result}
\langle P(t) P^*(0) \rangle = \frac{S}{\Gamma_X} \left\langle \exp{\left[-\mathrm i\omega_X t -\frac{\Gamma_X t}{2} - \rmi \beta
    \bar n_R t- \rmi \beta \int_0^{t}\delta n_R(t_1) \mathrm dt_1
  \right]}\right \rangle.
\end{equation}
Here the averaging over the random source realizations and reservoir population is
assumed. The fluctuations of the reservoir particle number
  $n_R$ take place on much longer timescale compared with the
  fluctuations of random forces, Eq.~(\ref{eq:random}). Hence, the
  averaging over the realizations of $f$ is carried out independently.

The oscillation spectrum, $\langle |P(\omega)|^2\rangle= \int_{-\infty}^\infty \langle P(t)P^*(0) \rangle \mathrm dt$,
which corresponds to the quantum dot emission spectrum in the weak coupling regime,
 depends
strongly on the relation between various timescales in the system:
exciton lifetime $1/\Gamma_X$, correlation time in the reservoir
$\tau_c$, defined by the relation $
 \langle \delta n_R(t)\delta n_R(0)\rangle=\langle \delta n_R^2\rangle\e^{-t/\tau_c}
$,
 and dephasing time $\tau_d$ caused by the reservoir, where
\begin{equation}
  \label{dephasing}
  \frac{1}{\tau_d} = \beta^2 \int_0^\infty \mathrm dt \, \langle \delta n_R(t)
    \delta n_R(0)\rangle =\beta^2\langle
    \delta n_R^2\rangle \tau_c.
\end{equation}
 If
reservoir fluctuations are fast and small enough, $\tau_c \ll \tau_d$,
$\Gamma_X^{-1}$, the so-called motional narrowing regime can be
realized and 
\begin{equation}
  \label{eq:fast}
  \langle |P^2(\omega)|\rangle \propto \frac{1}{(\omega -\omega_X -
    \beta \bar n_R)^2+
    [\Gamma_X/2+ 1/(2\tau_d)]^2}.
\end{equation}
In this case interactions with reservoir slightly broaden the quantum
dot spectrum. 

In the opposite case, where the fluctuations are strong and slow, the
quantum dot spectrum can be presented as 
\begin{equation}
  \label{eq:slow}
  \langle |P(\omega)|^2\rangle \propto {\int \mathrm d n_R\, p(n_R)\delta(\omega -\omega_X- \beta
    n_R)=\frac{1}{\beta}p\left(\frac{\omega-\omega_X}{\beta}\right),}
\end{equation}
where $p(n_R)$ is the distribution function of the reservoir, i.e. the
probability that the number of particles in the reservoir is $n_R$. 
The quantum dot spectrum Eq.~\eqref{eq:slow} in the regime of strong fluctuations is strongly asymmetric and reflects
the distribution of the particles in the reservoir.
Equation~(\ref{eq:slow}) is valid for the frequencies $|\omega-\omega_X - \beta \bar n_R| \gtrsim \Gamma_X$, that is why the spectral function of the quantum dot is replaced by $\delta$-function in Eq.~\eqref{eq:slow}. This equation also holds provided that the particle number fluctuations in the reservoir are so large that
\begin{equation}
\label{eq:large:fluct}
\beta \sqrt{\langle \delta n_R^2\rangle} \gg \Gamma_X, \tau_c^{-1}.
\end{equation}
In this case the dephasing takes place at a timescale $\sim (\beta\sqrt{\langle \delta n_R^2\rangle})^{-1}$ and it is not sensitive neither to the oscillator lifetime nor to the correlation time of the reservoir.

\begin{figure}
 \begin{center}
  \includegraphics[width=0.7\textwidth]{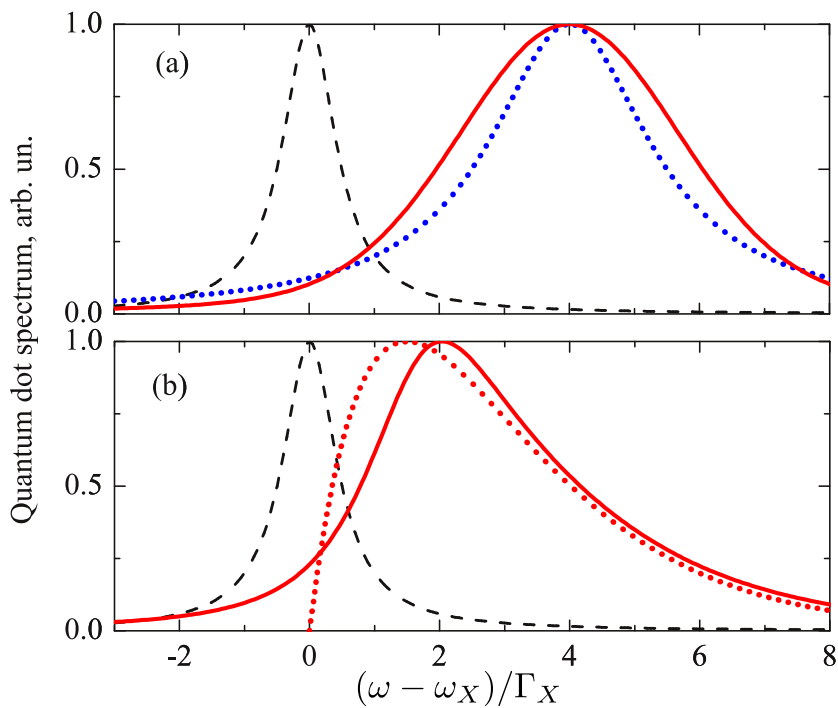}
 \end{center}
  \caption{(Color online) Effect of interaction with the reservoir (a) and within the quantum dot (b) on its spectrum $\langle |P(\omega)|^2\rangle$. Black/dashed, red/solid and blue/dotted curves in panel (a) correspond to linear regime, fast reservoir fluctuations regime and regime of slow reservoir fluctuations with Gaussian statistics, respectively.
Calculation was performed at $\beta \bar n_R=4\Gamma_{X}$, $1/\tau_{d}=2\Gamma_X$ and $\beta\delta\bar n_R=1.5\Gamma_X$. 
Black/dashed, red/solid and red/dotted curves in panel (b) correspond to linear regime, exact solution, Eq.~\eqref{eq:gen}, and approximate solution Eq.~\eqref{eq:highPump}, with $\delta$-function replaced by Lorentzian with full-width at half-maximum equal to $\Gamma_X$. Calculation was performed at $\alpha S=3\Gamma_{X}$.
}\label{fig:one_dot}
\end{figure}

These two limiting cases of reservoir fluctuations described by Eqs.~\eqref{eq:fast} and \eqref{eq:slow} are illustrated in Fig.~\ref{fig:one_dot}(a) where the calculated quantum dot spectra are shown. Black/dashed curve presents spectrum for vanishing reservoir fluctuations, this spectrum is described by the Lorentzian with the full width at half-maximum (FWHM) equal to $\Gamma_X$, centered at the resonance frequency $\omega_X$. Blue/dotted curve corresponds to the regime of fast reservoir fluctuations, Eq.~\eqref{eq:fast}. This spectrum is also a Lorentzian, shifted to larger energies and broadened  as compared to the linear regime. Red/solid curve is calculated in the regime of slow reservoir fluctuations. We have assumed, that the reservoir population is characterized by a Gaussian distribution with given mean value $\bar n_R$ and dispersion $\delta n_R$. 
In numerical calculations the $\delta$-function in Eq.~\eqref{eq:slow} was replaced by the Lorentzian with FWHM equal to $\Gamma_{X}$. Then the convolution \eqref{eq:slow} yields Voigt distribution with  Gaussian-like central part and Lorentzian wings, see Fig.~\ref{fig:one_dot}(a).  

It is noteworthy that for the quantum dot placed into the microcavity, the emission spectrum in the limit of weak and fast fluctuations, $\tau_c \ll \tau_d, \Gamma_X^{-1}$ is given by Eq.~\eqref{eq:linear} where $\omega_X$ is replaced by $\omega_x + \beta \bar n_R$ and $\Gamma_{X}$ is replaced by $\Gamma_X+ \tau_d^{-1}$. The qualitative shape of the spectra in the case of strong or slow reservoir fluctuations can be obtained by the convolution of the reservoir distribution function $p(n_R)$ with Eq.~\eqref{eq:linear} where $\omega_X$ is substituted by $\omega_X+ \beta n_R$. The resulting spectra are similar to those discussed in Sec.~\ref{sec:1osc_photon} where the case of the nonlinearity within the dot is addressed. 

\section{Interaction within the dot}\label{sec:nonlin}
Now we focus on the non-linear effects due to the interaction of the excitons within the same quantum dot. Hereinafter we disregard fluctuations of the particles in the reservoir studied above in Sec.~\ref{sec:reservoir}. Below we present one after another the studies of (i) the single dot case (Sec.~\ref{sec:1osc}), (ii) the single dot coupled with the cavity mode (Sec.~\ref{sec:1osc_photon}), and (iii) two quantum dots coupled with the cavity (Sec.~\ref{sec:2dots}).

\subsection{Single dot}\label{sec:1osc} 

The physical picture of the interaction effects within the dot on the emission spectra can be most transparently presented for the case of the single quantum dot which does not interact with the photon. Similarly to the situation studied in Sec.~\ref{sec:reservoir} its spectral function $\langle |P(\omega)|^2\rangle $ defines the emission spectrum in the weak coupling regime.

We start with the equation describing quantum dot as a non-linear oscillator:
\begin{equation}\label{eq:single}
 \frac{dP}{dt}=-\left(\rmi
\omega_{X}+\rmi\alpha|P|^2+\frac{\Gamma_{X}}{2}\right)P+f(t)\:,\quad \langle
f(t)f^*(t')\rangle =S\delta(t-t')\:.
\end{equation}
Similarly to the case of reservoir fluctuations, Eq.~\eqref{1:res}, the interaction term in
 Eq.~\eqref{eq:single}
leads to the blueshift and broadening of the oscillation spectrum. 
 Consequently, the analysis of reservoir fluctuations, performed above, may be used here. 
The strength of the fluctuating non-linear term is determined by the value of pumping.
Large pumping ($\alpha \langle |P|^2\rangle \gg \Gamma_X$) corresponds to the regime of strong fluctuations, cf. Eq.~\eqref{eq:large:fluct}. Hence, the quantum dot spectrum may be presented in the form, similar to Eq.~\eqref{eq:slow},
\begin{equation}\label{eq:stochlin}
 \langle |P(\omega)|^2\rangle=\int\rmd \Omega
F(\Omega)\delta(\Omega-\omega)\:,
\end{equation}
where the distribution function $F(\Omega)$ with the oscillator frequency $\Omega = \omega_X + \alpha |P|^2$, is determined from the statistics of the non-linear term in Eq.~\eqref{eq:single}.

The major difference of the non-linear Eq.~\eqref{eq:single} and the linear one Eq.~\eqref{1:res} describing the interaction of the quantum dot exciton with the reservoir is that the fluctuations of the polarization $P$ themselves govern the blueshift and, on the other hand, the blueshift determines the fluctuations of $P$. In order to determine the distribution function $F(\Omega)$ we use the stochastic linearization of Eq.~\eqref{eq:single} described in detail in Refs.~\cite{budgor1976,Spanos2011}. The starting point of the stochastic linearization is the time-dependent Eq.~\eqref{eq:single} where decay and pumping terms are neglected, $\Gamma_X \equiv 0$, $f(t) \equiv 0$. The solution of Eq.~\eqref{eq:single} is then given by $P(t)=|P|\rme^{-\rmi\Omega t}$, where
\begin{equation}\label{eq:to_linearize}
\Omega|P|^2=\omega_X|P|^2+\alpha |P|^4\:.
\end{equation}
In the presence of the decay and pumping Eq.~\eqref{eq:to_linearize} does not hold. In the stochastic linearization approach the difference of the left- and right- hand sides of Eq.~\eqref{eq:to_linearize} should be minimized. It implies, in particular, that  the average (over the random sources realizations) blueshift of the resonance frequency is given by
\begin{equation}\label{eq:Omega}
\langle \Omega-\omega_X\rangle=\frac{\alpha\langle|P|^4\rangle}{\langle |P|^2\rangle}\:.
\end{equation}
Hereinafter we use the notation
$\langle |P|^2\rangle\equiv\langle |P(t=0)|^2\rangle=\int \rmd \omega \langle|P(\omega)^2|\rangle/(2\pi)$.
Equation~\eqref{eq:Omega} is automatically satisfied when the distribution of the frequency $\Omega$ is chosen in the form 
\begin{equation}
 F(\Omega)=\mathcal N(\Omega-\omega_X)p\left(\frac{\Omega-\omega_X}{\alpha}\right)\:, 
\label{eq:Omega:distrib}
\end{equation}
where  $\mathcal N$ is the normalization constant,
\begin{equation}
 p(|P|^2)=\frac1{\langle|P|^2 \rangle}\exp\left(-\frac{|P|^2}{\langle|P|^2 \rangle}
\right)\:,\label{eq:distr}
\end{equation}
is the distribution function of the exciton polarization absolute value, and
\begin{equation}
 \langle |P^2| \rangle=\frac{S}{\Gamma_{X}}\label{eq:avP}\:,
\end{equation}
is given by the ratio of the pumping  and decay rates. The shape of this distribution is independent of the pumping rate. In other words, the non-linear term in Eq.~\eqref{eq:single} influences the oscillator phase only and does not affect its amplitude. Thus, the distribution Eq.~\eqref{eq:distr} is the same as in the linear regime and it inherits the Gaussian statistics of the noise term fluctuations.  Using Eq.~\eqref{eq:Omega} and Eq.~\eqref{eq:distr} we find that
\begin{equation}\label{eq:highPump}
 \langle |P(\omega)|^2\rangle\propto F(\omega)=\cases{
(\omega-\omega_X)\exp\left(-\frac{\omega-\omega_X}{\alpha \langle |P^2|
\rangle}\right),&$\omega>\omega_X$\cr
0, &$\omega<\omega_X$.}
\end{equation}
This function decays exponentially for large values of $\Omega$  because the  realization of  high blueshift $\alpha |P^2|$ is unlikely. Moreover, frequencies $\omega<\omega_X$ are not possible since interactions are repulsive and lead to the increase of energy only. Equation~\eqref{eq:highPump} demonstrates, that in the regime of high pumping the oscillator spectrum is strongly asymmetric and broadened due to the fluctuations of the resonance frequency.

For arbitrary value of pumping strength the general analytical result for the oscillator spectrum can be obtained  by means of  the Fokker-Planck equation technique~\cite{Risken_1989,Poddubny2010prb}, similarly to the case of the noise-driven Duffing oscillator~\cite{Dykman1980}. The spectrum reads
\begin{equation}\label{eq:gen}
 \langle |P(\omega)^2|\rangle\propto
\Im\sum\limits_{n=0}^\infty\frac{(I_n)^2}{\lambda_{n}-\omega}\:,
\end{equation}
where
\begin{equation*}
  \lambda_{n}=\omega_X+\rmi\frac{\Gamma_X}{2}-2\rmi (n+1)
\sqrt{\left(\frac{\Gamma_X}{2}\right)^2+\rmi\alpha S}\:,
\end{equation*}
\begin{equation*}\label{eq:In}
 I_n=\frac{4b\sqrt{(n+1) a}}{(a+b)^2} \left(\frac{a-b}{a+b}
\right)^n, \quad
 a=\frac{\Gamma_{X}}{S},\quad b=\sqrt{a^2+\frac{4\rmi  \alpha}{S}}\:.
\end{equation*} 
In the linear in the pumping regime where $S\to 0$ only the first term in the series Eq.~\eref{eq:gen} does not
vanish and the result is reduced to Lorentzian with FHWM equal to $\Gamma_X$. For very large pumping,
$\alpha \langle |P^2|\rangle\gg\Gamma_{X}$, the series reduce
to Eq.~\eqref{eq:highPump}.

The spectra of the single oscillator, $\langle |P(\omega)|^2\rangle$, calculated for different  pumping strengths, are shown in Fig.~\ref{fig:one_dot}(b). The spectrum at the strong pumping represented by red/solid curve, is strongly broadened and shifted as compared to the spectrum found in the linear regime. The latter is shown by black/dashed black curve.   The  width of the non-linear spectrum is of the same order as the blueshift. The spectrum found by the stochastic linearization, Eq.~\eref{eq:highPump}, and shown by red/dotted curve in Fig.~\ref{fig:one_dot}(b) well approximates the exactly calculated one, Eq.~\eqref{eq:gen}. The strong asymmetry of emission spectra calculated for the non-linear regime is clearly seen from the Figure.

This concludes the discussion of the single oscillator and now we proceed to the analysis of the quantum dot coupled with the cavity mode.


\subsection{Single dot coupled with the cavity}\label{sec:1osc_photon}

To start, we recall that in the linear-in pumping regime,
$S\to 0$, 
and under the conditions of the strong coupling, $g\gg \Gamma_C,\Gamma_X$, the spectrum described by Eq.~\eqref{eq:PLlin} with $N=1$ consists of the two distinct peaks at the frequencies of the system eigenmodes, excitonic polaritons. The main question we address here is how this two-peak spectrum changes with allowance for the non-linearity. In the general case the emission spectrum under strong pumping should be calculated numerically. This can
be done either reducing the problem to the Fokker-Planck equation
\cite{Risken_1989} or directly integrating the set of Eqs.~\eref{eq:main}. The latter procedure turns out to be more efficient, because the Fokker-Planck equation  is computationally demanding already for $N=1$ due to the large number of independent variables. In our calculations we have used the simplest 1.5-order Heun method for  integration of stochastic differential equations~\cite{kloeden2010,burrage2007}.

\begin{figure}
 \begin{center}
  \includegraphics[width=0.7\textwidth]{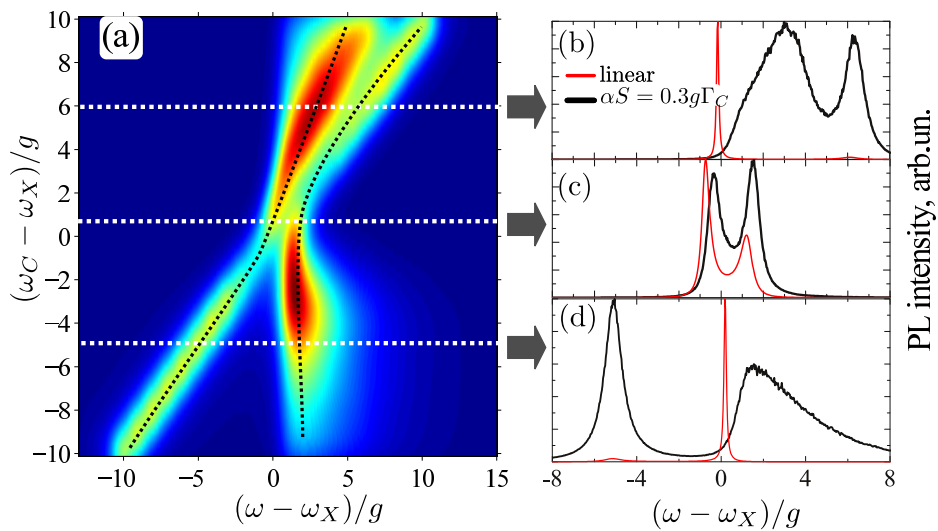}
 \end{center}
  \caption{(Color online) (a) Photoluminescence spectra for single quantum dot coupled  to microcavity mode as function of the detuning between cavity and exciton frequencies. Panels (b),(c),(d) show the spectra for $\omega_C=\omega_X+6g$, $\omega_C=\omega_X+0.5g$, $\omega_C=\omega_X-5g$, respectively. Black/thick  and red/thin lines in panels (b), (c), and (d) correspond to non-linear and
linear in pumping regime, $\alpha S=0.3g\Gamma_{C}$ and $S=0$, respectively. These curves are normalized to their maximum values.
Black/dashed curves in panel (a) are guides for eye demonstrating anticrossing. Calculations are carried out at $\Gamma_{X}=0.1g$, $\Gamma_C=g$, $\alpha S=0.3g\Gamma_{C}$. The time-domain integration was performed up to $t=100/g$, with the time step $0.003/g$,  the spectra were averaged over ensemble
with $4000$ realizations of random forces.
}\label{fig:PL1}
\end{figure}

Luminescence spectra calculated for different detunings between the exciton and photon modes are presented in Fig.~\ref{fig:PL1}. Panel (a) shows the color plot of the emission intensity. Panels (b), (c), and (d) present the spectra for the detuning $\omega_C-\omega_X$ equal to $6g$, $0.5g$ and $-5g$, respectively. 
Thin/red and thick/black curves are calculated, respectively, for (i) the linear regime $S\to 0$ and (ii) the regime where the non-linearity is already  strong, $S=0.3\Gamma_C/\alpha$. The spectra are normalized to their maximum values, the other parameters of the calculations are given in the caption to the Figure.
The calculation demonstrates, that the non-linear spectra retain the characteristic two-peak structure, although 
the spectral shape is strongly affected by the non-linearity, in particular, it becomes asymmetric, as well seen in Figs.~\ref{fig:PL1}(b) and (d). The spectral maxima in the non-linear regime clearly exhibit the
anticrossing behavior, see Fig.~\ref{fig:PL1}(a). From this we conclude that
the strong coupling survives even for non-linear regime if $\alpha\langle |P^2|\rangle\lesssim g$.
\begin{figure}
 \begin{center}
  \includegraphics[width=0.5\textwidth]{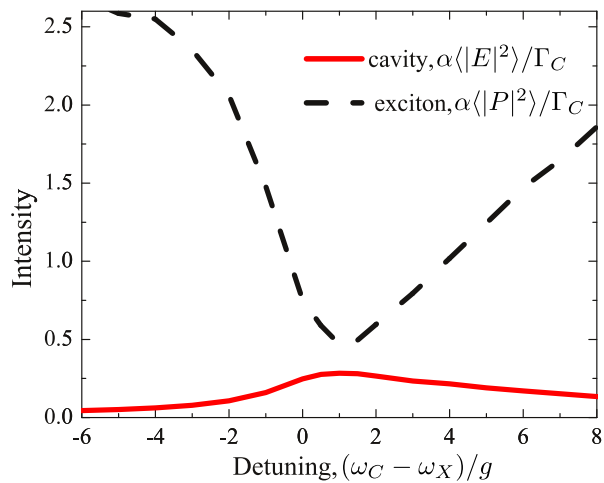}
 \end{center}
  \caption{(Color online)  Intensities of photon and exciton modes
as functions of the detuning $\omega_C-\omega_X$.
Red/solid and black/dashed curves correspond to $\alpha\langle |E|^2\rangle/\Gamma_C$ and $\alpha \langle |P|^2\rangle/\Gamma_C$, respectively.
The parameters of calculations are the same as in Fig.~\ref{fig:PL1}. 
}\label{fig:population}
\end{figure}

Let us now discuss the spectra in Fig.~\ref{fig:PL1} in more detail.
We start from the case of large detuning between exciton and photon modes, $|\omega_X - \omega_C|\gg g$. In this situation the spectrum has two peaks related with the cavity and exciton emission. It is noteworthy, that 
 in the non-linear regime, the blueshift of the exciton is determined by the dot population which, in its turn, is related with the detuning: $\langle |P|^2\rangle \approx S(\omega_X - \omega_C)^2/(\Gamma_C g^2)$ (provided that $\Gamma_X \ll \Gamma_C$ and $\Gamma_C \ll |\omega_X - \omega_C|$). Hence, if bare exciton frequency $\omega_X$ is fixed, but the cavity frequency $\omega_C$ is varied, the position of exciton peak changes due to the variation of the blueshift.

 To confirm this argument we have plotted in Fig.~\ref{fig:population} the dependence of the stationary intensities $\langle |E|^2\rangle$ and $\langle |P^2|\rangle$ on the detuning. For large detuning one has $\langle |P^2|\rangle\gg \langle|E|^2\rangle$ which is explained by the longer exciton lifetime as compared to the photon lifetime, $\Gamma_C\gg \Gamma_X$.  For small values of detuning the curves become closer to each other as a result of the coupling between exciton and photon modes.

Another important feature revealed in Fig.~\ref{fig:PL1} is the strong asymmetry
between the spectral shapes of the exciton peak for large positive and negative
detuning, cf. Fig.~\ref{fig:PL1}(b) and 
Fig.~\ref{fig:PL1}(c). 
This is quite different from the linear regime, where the spectra
Eq.~\eqref{eq:PLlin} are symmetric with respect to zero detuning.
In the non-linear case the exciton peak is asymmetrically broadened due to the frequency fluctuations, 
like in the case of single oscillator, and the shape of
the broadened peak depends on the sign of detuning. For negative
detuning, $\omega_C-\omega_X<0$, the high-energy tail dominates the spectrum similarly to the case of the dot decoupled from the cavity, see Sec.~\ref{sec:1osc} and Fig.~\ref{fig:one_dot}(b), while for
positive detuning, $\omega_C-\omega_X>0$, this tail is quenched. This is related with the fact that strong fluctuations of exciton polarization $P$ are strongly suppressed since they correspond to large blueshifts where exciton mode approaches the cavity mode and, hence, short lifetimes.

\begin{figure}
 \begin{center}
  \includegraphics[width=0.7\textwidth]{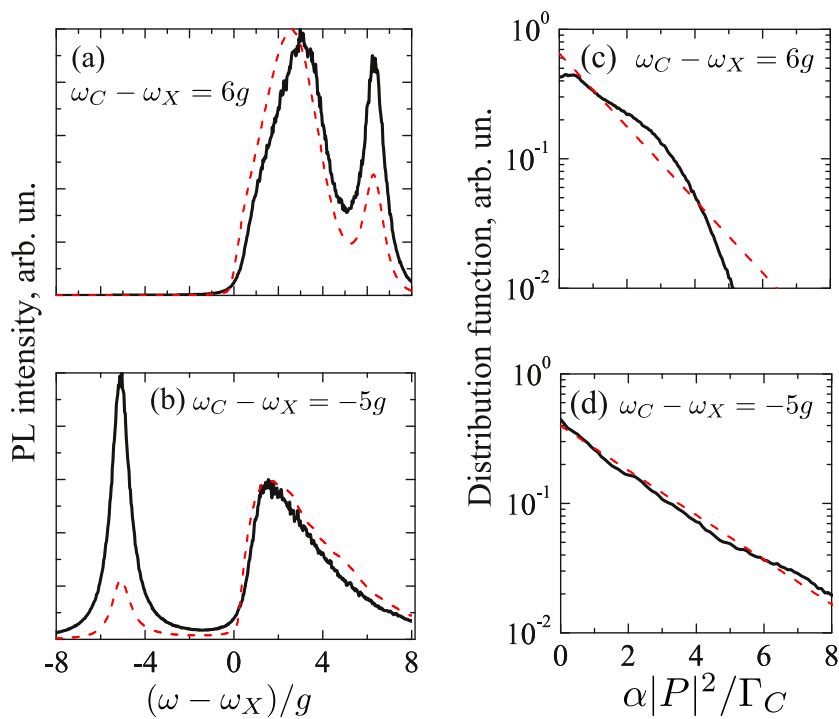}
 \end{center}
  \caption{(a), (b) Cavity emission spectra calculated for the detuning $\omega_{C}-\omega_{X}=6g$ (a),
and $\omega_{C}-\omega_X=-5g$ (b). (c), (d) Distribution functions of the intensity $ |P|^2$ calculated for the same parameters as in panels (a) and (b), respectively. Black/solid  lines represent results of direct numerical calculation, red/dashed  lines are calculated according to Eqs.~\eqref{eq:distr},\eqref{eq:convolution}.
Other parameters of calculations  are the same as in Fig.~\ref{fig:PL1}. 
}\label{fig:understand}
\end{figure}

To understand this effect in depth we have plotted in Fig.~\ref{fig:understand} the emission spectra [(a), (b)] along with the distribution functions  of the quantum dot exciton intensity $|P|^2$. Distribution functions $p(|P|^2)$, shown by black/solid curves in Fig.~\ref{fig:understand}(c) and Fig.~\ref{fig:understand}(d), were extracted from the numerical solutions of the system \eqref{eq:PLlin}. For comparison, red/dashed curves show exponential distributions \eqref{eq:distr} with the same average values of $|P|^2$.  
The emission spectrum of the microcavity can be presented in the following phenomenological form similar to Eq.~\eqref{eq:slow} and Eq.~\eqref{eq:stochlin} obtained in the stochastic linearization method:
\begin{equation}\label{eq:convolution}
 I(\omega) \propto\int \rmd \Omega F(\Omega) I_{\rm
lin}(\omega)\:.
\end{equation}
where $I_{\rm lin}(\omega)$ is given by Eq.~\eqref{eq:PLlin} with $\omega_X = \Omega$ and the distributions $F(\Omega)$ and $p(|P^2|)$ are related by Eq.~\eqref{eq:Omega}.
Corresponding spectra, calculated with numerically found functions $p(|P|^2)$, are shown in Fig.~\ref{fig:understand}(a) and Fig.~\ref{fig:understand}(b) by red/thin curves.  We see, that Eq.~\eqref{eq:convolution} satisfactory describes the shape of the exciton peaks for both signs of the detuning and clearly demonstrates the correspondence between the shape of the exciton peak and the distribution function of the intensity $|P|^2$.
For negative detuning the distribution function and the excitonic peak have exponential tails, similarly to the case of single oscillator, Eq.~\eqref{eq:highPump}. Suppression of this tail for positive detuning means that the distribution function decays faster than exponential [cf. red and black curves in Fig.~\ref{fig:understand}(c)] which is explained by interaction of excitonic and photonic modes. Indeed, for $\omega_C>\omega_X$ the non-linear blueshift of the exciton frequency decreases the detuning. Consequently,  the exciton lifetime becomes effectively smaller due to the Purcell effect which suppresses   the probability of such large detuning. Another effect leading to the same result is the repulsion of the excitonic mode with high blueshift from the cavity mode. In the opposite case,  $\omega_C<\omega_X$, the absolute value of the detuning is further increased by the blueshift, so  the shape of the excitonic peak is not modified by the interaction with the cavity.

Phenomenological Eq.~\eqref{eq:convolution} fails, however, to reproduce the ratio between the magnitudes of the excitonic and photonic peaks in the spectra. Formally, Eq.~\eqref{eq:convolution} is valid provided that the timescale of the fluctuations of $P$ is large compared with other timescales in the system [cf. Eq.~\eqref{eq:large:fluct}] or the magnitude of the blueshift exceeds by far all other energy scales. Neither of these conditions holds in the system under study.

\begin{figure}[t]
 \begin{center}
  \includegraphics[width=0.7\textwidth]{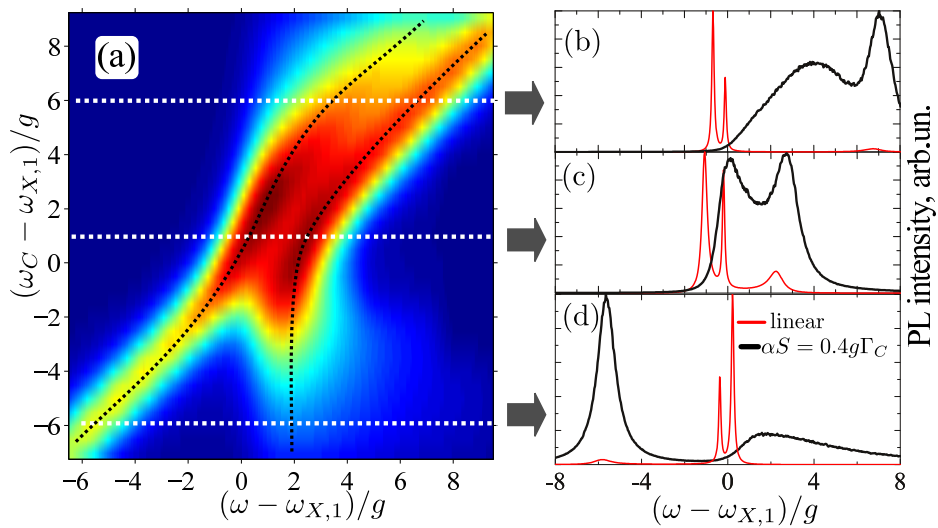}
 \end{center}
  \caption{
 (a) Photoluminescence spectra for two quantum dots coupled  to microcavity
mode as function of the detuning between cavity and exciton dot.
Panels (b), (c), (d) show the spectra for $\omega_C=\omega_{X,1}+6g$,
$\omega_C=\omega_X+g$, $\omega_C=\omega_X-6g$, respectively. 
Black/thick and red/thin lines in panels (b) -- (d) correspond to non-linear and linear in pumping regime, $\alpha S=0.4g\Gamma_{C}$ and $S\to0$, respectively. These curves are normalized at their maximum values. 
Black/dashed curves in panel (a) are guides for eye demonstrating anticrossing. 
Calculation was carried out for
$\omega_{X,1}-\omega_{X,2}=g$, other parameters are the same as in Fig.~\ref{fig:PL1}.
}\label{fig:PL2}
\end{figure}

\subsection{Two dots coupled with the cavity}\label{sec:2dots}
Now we turn to the discussion of the emission spectra of the cavity with two embedded dots, shown in Fig.~\ref{fig:PL2}. The specific feature of this problem in the linear regime is the formation of the collective, superradiant mode of the quantum dot excitons~\cite{keldysh07,JETP2009,Poddubny2010prb}. In particular, as discussed above in Sec.~\ref{sec:model}, there are three eigenmodes of the homogeneous linear system Eq.~\eqref{eq:main} for $N=2$. In case where $\omega_{X,i}=\omega_{X,2}$, one of these modes, with the excitons oscillating with the opposite phases, $P_{1}=-P_{2}$, does not interact with the cavity mode, it is called dark. Two remaining modes correspond to the excitons, oscillating in phase, 
$P_{1}=P_{2}$, and are formed by the coupling between the superradiant mode of the excitons and the cavity mode.  Only two peaks are manifested in the emission spectrum. The superradiant effect leads to the enhancement of the Rabi splitting between these peaks from $g$ to $\sqrt{N}g$, see Eq.~\eqref{eq:PLlin}. When the frequencies of the dots $\omega_{X,i}$ are different, or the exciton tunneling between the dots is introduced, the dark modes mix with the superradiant mode and the spectrum acquires three-peak shape. As is demonstrated in previous  work \cite{JETP2009}, the superradiant mode is stable against the disorder when the characteristic splitting of the dot frequencies is less than $g\sqrt{N}$. Here we analyze the stability of the superradiant mode against the interactions.

Figure~\ref{fig:PL2} shows the emission spectra of the microcavity containing two quantum dots for various frequencies of the cavity mode. The frequency separation between the dots was fixed to be $\omega_{X,2}-\omega_{X,1}=g$ and the $\omega_C$ was varied. Overall intensity dependence on the emission and cavity frequency shown in Fig.~\ref{fig:PL2}(a) is quite similar to the intensity distribution for one dot in the microcavity [see Fig.~\ref{fig:PL1}(a)] and clearly demonstrates two peaks for any given cavity mode position. Thus, the superradiant mode is stable against the interaction.

Panels (b), (c), and (d) of Fig.~\ref{fig:PL2} present the details of emission spectra for three different cavity mode positions. Black/solid curves are calculated for the strong pumping where the nonlinear behavior is pronounced, red/thin curves correspond to the linear regime. The emission spectra in the non-linear regime corresponding to rather large detunings [panels (b), (d)] demonstrate the asymmetry studied for the single dot in the cavity in Sec.~\ref{sec:1osc_photon}. Interestingly, the spectra in the linear regime demonstrate three peaks: two stronger ones correspond to the dot emission and weaker one to the cavity emission. The presence of non-linearity induced by the strong pumping qualitatively changes the emission spectrum: excitonic peaks merge and their intensity drops. Note, however, that the strong coupling regime is still maintained, see anticrossing in Fig.~\ref{fig:PL2}. The smallest splitting between the maxima of the  spectra is larger than the Rabi splitting for the single dot $2g$. This is a fingerprint of the superradiant mode stability in the non-linear regime.

Thus, the superradiant mode becomes stabilized by interactions. Indeed, for positive detunings, $\omega_C>\omega_{X,1}, \omega_{X,2}$, the blueshifts of the quantum dots are different: The dot with higher exciton frequency (dot 2 in our calculation) has smaller lifetime due to proximity to the cavity mode, correspondingly, smaller exciton intensity $\langle |P_2|^2\rangle<\langle|P_1|^2\rangle$, and, hence, smaller blueshift. Hence, with an increase of the pumping rate the blueshifted dot frequencies approach to each other, resulting in the decrease of the splitting between exciton frequencies and stabilization of the superradiant mode.

\section{Conclusions}\label{sec:concl}

To conclude, we have developed a theory of non-linear emission of quantum dots coupled to the optical mode of the microcavity under non-resonant excitation. We model quantum dot excitons as non-linear oscillators taking into account the repulsive exciton-exciton interactions both within the dot and between the quantum dot and excitonic reservoir. We use random sources approach to model the relaxation processes and apply stochastic linearization and numerical integration of the Langevin equations to determine the spectra.  

Our model clearly shows that interactions: (i) blueshift the transition energy and (ii) to the same extent broaden the emission peak. The interactions result in the complex behavior of the exciton lifetimes and intensities of emission as functions of pumping rate. The emission lines are strongly asymmetric and bear information on the exciton statistics. Contrary to the linear regime, the lineshapes are sensitive to the sign of the detuning between the exciton and photon modes. Interestingly, even for substantial pumping the strong coupling regime between the cavity mode and the quantum dot exciton can be preserved. Moreover, if two quantum dots are placed in the cavity, the superradiant behavior can be stabilized by the pumping.

\ack

Financial support of RFBR,  RF President Grant NSh-5442.2012.2, and EU project POLAPHEN is
gratefully acknowledged.


\providecommand{\newblock}{}


\end{document}